# Comparative Robustness of Mach–Zehnder and Microring Photonic Matrix Multipliers Under Laser RIN and Thermal Drift

**Andreas Falkenberg**
QPI Laboratories, Escondido, CA, USA
andreas@qpilabs.com

**Abstract.** Photonic matrix multipliers can accelerate neural-network inference, but their practical utility depends on how optical weight elements respond to source noise, thermal variation, and calibration error. This paper compares Mach–Zehnder modulator (MZM) and microring-resonator (MRR) implementations under laser relative-intensity noise (RIN), column-wise thermal/phase drift, and their joint action. Two CIFAR-10 classifiers—a compact four-layer CNN and ResNet-8—are mapped to matched photonic multiply–accumulate chains using common source-distribution, detection, numerical, and calibration assumptions. Both architectures reproduce the same zero-perturbation baseline for each workload: 79.5% classification accuracy for the Small CNN and 70.0% for ResNet-8. Robustness is then evaluated using separate RIN-only and drift-only sweeps and a full two-dimensional RIN–drift grid, with dimensionless RMS perturbation levels extending from 0.5% to 20%. The analysis reports classification accuracy, class-normalized confusion matrices, rolling-accuracy traces, one-dimensional robustness curves, and multidimensional response surfaces. Device equations and parameter definitions follow the models implemented in the QPI Labs Photonic Inference Simulator, with shared laser fluctuations propagated as correlated multiplicative noise and slow drift represented as a per-output-column perturbation of calibrated weights. All numerical results are generated from recorded experiment configurations and random seeds. The study provides a controlled system-level framework for comparing interferometric and resonant photonic weight elements while limiting conclusions to the implemented device models, perturbation ranges, calibration policy, and neural-network workloads.

## I. Introduction

Analog photonic processors exploit optical propagation and detection to evaluate linear transforms with high parallelism [7], [8]. Their value for neural-network inference is determined not only by nominal compute density, but also by the extent to which optical nonidealities alter network-level decisions. This study therefore compares the two implemented weight-element models using matched neural workloads, source and distribution conditions, photodetection, calibration, and numerical mapping. The reported differences are specific to these simulated configurations and perturbation models.

This study evaluates two implementations of the same photonic matrix-multiplication function. The MZM implementation uses an interferometric weight response, whereas the MRR implementation uses a resonant weight response. In the evaluated system model, slow thermal and phase variation is represented as a per-output-column multiplicative perturbation of the calibrated weights for both architectures. Both implementations are evaluated under the same configured source, distribution, detection, and numerical conditions. Silicon's thermo-optic coefficient is approximately $1.86 \times 10^{-4}\ \mathrm{K}^{-1}$, and published silicon-ring studies report temperature-dependent resonance shifts on the order of tens of picometers per kelvin [1], [2]. These device characteristics motivate a direct comparison of RIN and thermal sensitivity at the inference level.

This study presents simulated inference results for MZM- and MRR-based photonic matrix multipliers under separate and joint laser RIN and thermal-drift perturbations. The comparison uses matched optical and numerical conditions and evaluates a compact CIFAR-10 CNN and ResNet-8 using classification accuracy. The architecture terminology and system-level classification follow the QPI Labs photonic-inference framework [12].

## II. Neural-Network Workloads

### A. CIFAR-10 Small CNN [4], [5]

The CIFAR-10 Small CNN [4], [5] contains two convolutional layers and two fully connected layers, with 268,650 trainable parameters and a zero-perturbation baseline test accuracy of 79.5%. Because the calibrated MZM and MRR mappings reproduce the same result when $\sigma_{\mathrm{RIN}} = \sigma_{\mathrm{drift}} = 0$, this value serves as the common baseline for both modulators. The network is mapped to four chained photonic multiply–accumulate (MAC) layers that execute 1.89 million MACs per inference. Fig. 1 provides a compact visual overview of the data flow, while Table I reports the exact tensor dimensions, activation functions, layer parameters, and MAC mapping required for reproducibility. The first convolution maps the $3 \times 32^2$ input to $16 \times 32^2$ features; after spatial reduction to $16 \times 16$, the second convolution produces $32 \times 16^2$ features. The classifier then maps 2,048 features to 128 hidden units and finally to 10 output classes. Each enabled nonideality is applied independently at every mapped layer, and the resulting layer outputs are propagated through the four-layer photonic chain.

TABLE I. CIFAR-10 SMALL-CNN ARCHITECTURE AND PHOTONIC MAC MAPPING

| Layer | Input | Output | Activation | Trainable parameters | Photonic mapping |
|---|---|---|---|---|---|
| **Conv1** | 3 × 32 × 32 | 16 × 32 × 32 | ReLU | 448 | MAC layer 1 |
| **Conv2** | 16 × 16 × 16 | 32 × 16 × 16 | ReLU | 4,640 | MAC layer 2 |
| **FC1** | 2,048 | 128 | ReLU | 262,272 | MAC layer 3 |
| **FC2** | 128 | 10 | Softmax | 1,290 | MAC layer 4 |
| **Total** | — | 10 classes | — | **268,650** | **1.89 million MACs/inference** |

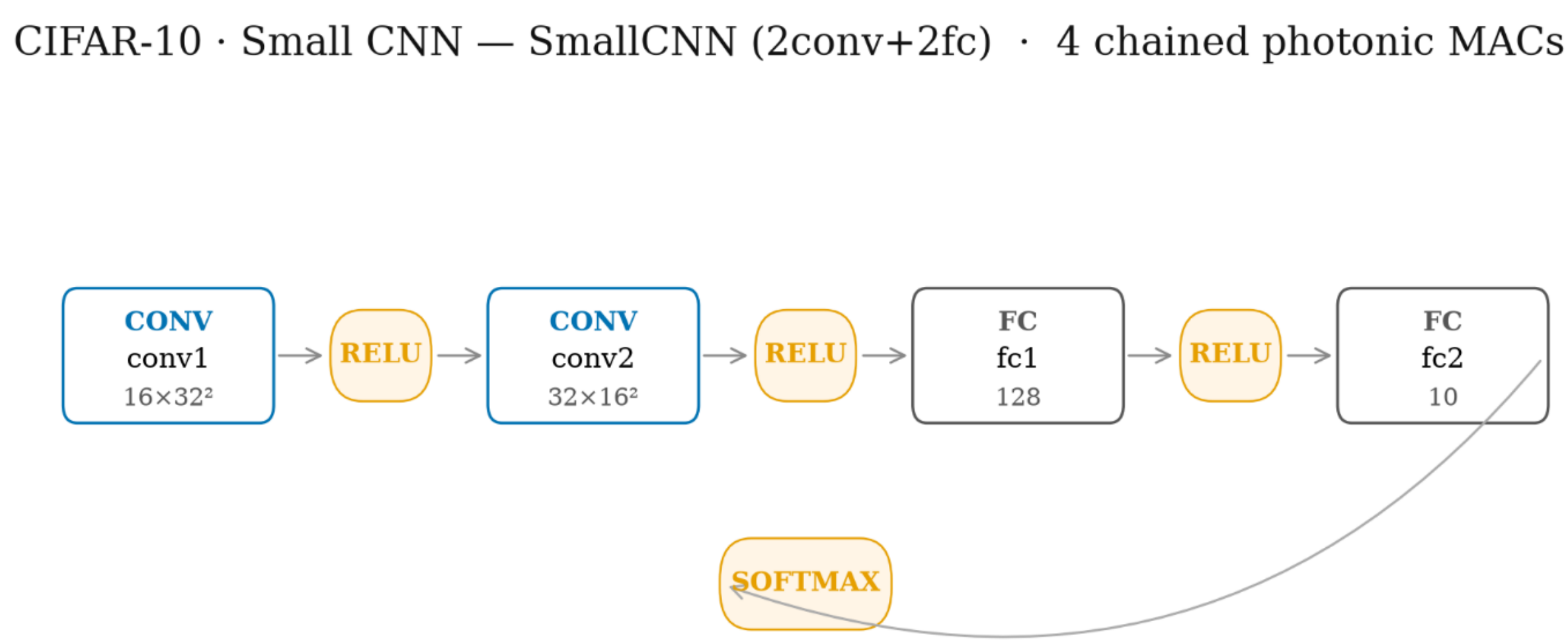

**Fig. 1. CIFAR-10 Small-CNN architecture and its mapping to four chained photonic MAC layers.**

### B. CIFAR-10 ResNet-8

The CIFAR-10 ResNet-8 workload [4], [6] contains a convolutional stem, three residual blocks, and a fully connected classifier, with 77,706 trainable parameters and a zero-perturbation baseline test accuracy of 70.0%. Because the calibrated MZM and MRR mappings reproduce the same result when $\sigma_{\mathrm{RIN}} = \sigma_{\mathrm{drift}} = 0$, this value serves as the common baseline for both modulators. The network is mapped to 10 chained photonic multiply–accumulate (MAC) layers that execute 12.50 million MACs per inference. Fig. 2 provides a compact visual overview of the residual data flow, while Table II reports the tensor dimensions, activation functions, layer parameters, and complete MAC mapping required for reproducibility. The photonic chain comprises the convolutional stem, two convolutions in each residual block, the projection shortcuts in blocks 2 and 3, and the final fully connected classifier. Each enabled nonideality is applied independently at every mapped layer, and the resulting layer outputs are propagated through the 10-layer photonic chain.

**TABLE II. CIFAR-10 RESNET-8 ARCHITECTURE AND PHOTONIC MAC MAPPING**

| Layer | Input | Output | Activation | Trainable parameters | Photonic mapping |
|---|---|---|---|---|---|
| **Stem Conv** | 3 × 32 × 32 | 16 × 32 × 32 | ReLU | 448 | MAC layer 1 |
| **Block 1, Conv1** | 16 × 32 × 32 | 16 × 32 × 32 | ReLU | 2,320 | MAC layer 2 |
| **Block 1, Conv2** | 16 × 32 × 32 | 16 × 32 × 32 | ReLU after residual addition | 2,320 | MAC layer 3 |
| **Block 2, Conv1** | 16 × 32 × 32 | 32 × 16 × 16 | ReLU | 4,640 | MAC layer 4 |
| **Block 2, Conv2** | 32 × 16 × 16 | 32 × 16 × 16 | ReLU after residual addition | 9,248 | MAC layer 5 |
| **Block 2, Projection** | 16 × 32 × 32 | 32 × 16 × 16 | None | 544 | MAC layer 6 |
| **Block 3, Conv1** | 32 × 16 × 16 | 64 × 8 × 8 | ReLU | 18,496 | MAC layer 7 |
| **Block 3, Conv2** | 64 × 8 × 8 | 64 × 8 × 8 | ReLU after residual addition | 36,928 | MAC layer 8 |
| **Block 3, Projection** | 32 × 16 × 16 | 64 × 8 × 8 | None | 2,112 | MAC layer 9 |
| **Global Avg. Pool + FC** | 64 | 10 | Softmax | 650 | MAC layer 10 |
| **Total** | — | 10 classes | — | **77,706** | **12.50 million MACs/inference** |

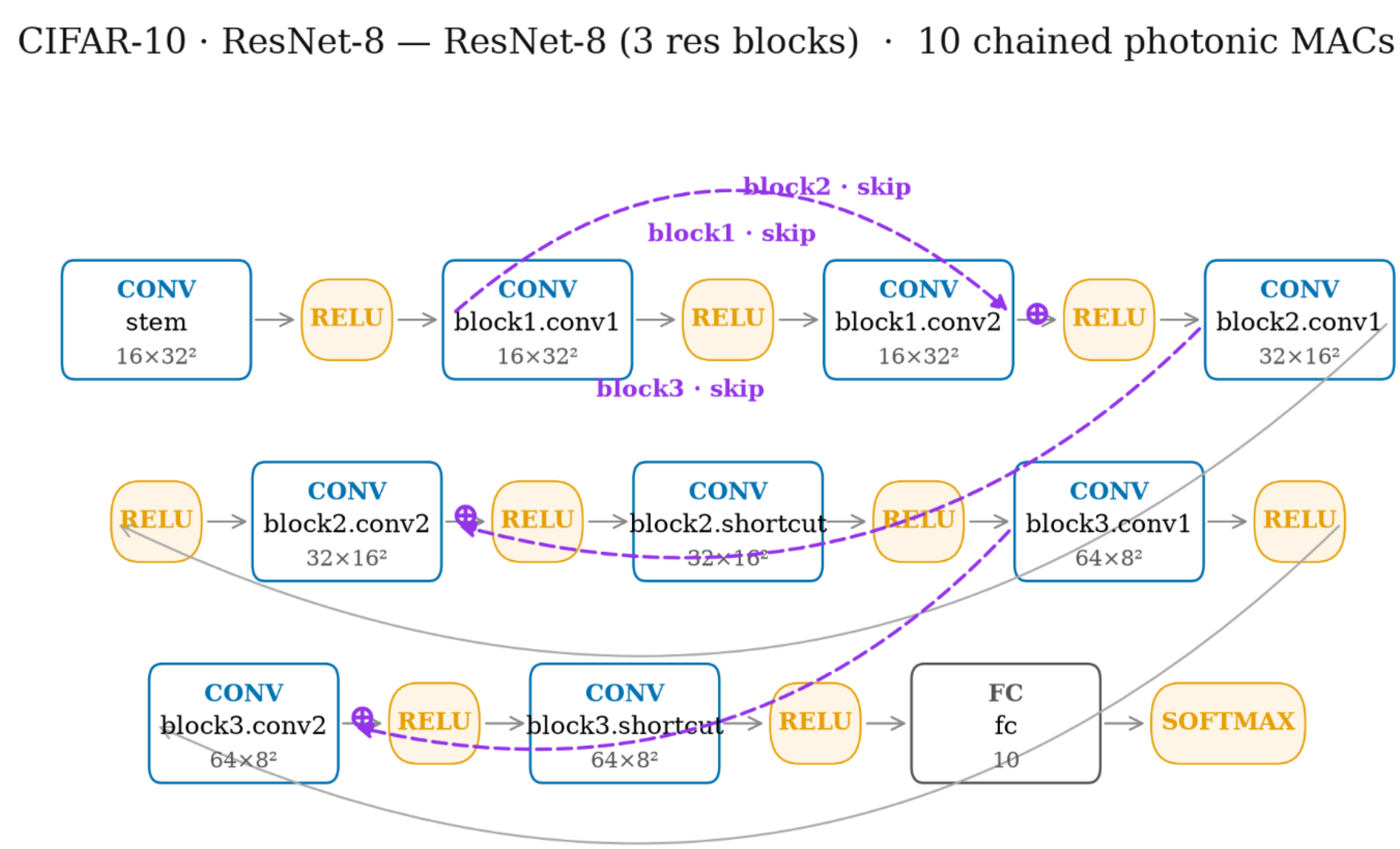

**Fig. 2. CIFAR-10 ResNet-8 architecture and its mapping to 10 chained photonic MAC layers.**

## III. Photonic Matrix-Multiplier Architectures

### A. Common Source, Fanout, and Detection

A common optical source and fanout network supplies activation-encoded optical power to the programmable weight elements. For source channel *i*, the power delivered to branch *m* is modeled as:

$$P_{i,m} = P_{s,i} x_i \eta_{f,m} = P_{s,i} x_i \rho_m 10^{-L_{f,m}/10}$$

where $P_{s,i}$ is the source power, $x_i$ is the normalized activation encoding, $\rho_m$ is the ideal power-split fraction assigned to branch *m*, $L_{f,m}$ is its excess and distribution loss in decibels, and $\eta_{f,m}$ is the corresponding net fanout efficiency. For an ideal equal *M*-way split, $\rho_m = \frac{1}{M}$ ; measured or configured imbalance is represented by branch-dependent $\rho_m$ and $L_{f,m}$.

Because the contributing optical channels are mutually incoherent in this study, their time-averaged detected powers add without coherent interference cross terms. After weighting, the optical power accumulated at output *j* is therefore

$$P_j^{\mathrm{out}} = \sum_{i=1}^{N} P_{i,m(j)} T_{ij}$$

where $T_{ij}$ is the programmed optical transmission representing matrix element $w_{ij}$, and *m*(*j*) identifies the fanout branch feeding output *j*. If signed weights are implemented by differential positive and negative paths, the detected signal is formed from their power difference, $\Delta P_j = P_{j,+}^{\mathrm{out}} - P_{j,-}^{\mathrm{out}}$.

The photodetector converts the accumulated optical power into current according to

$$I_j = R_{\mathrm{pd}} P_j^{\mathrm{out}} + n_j$$

where $R_{\mathrm{pd}}$ is the detector responsivity and $n_j$ is the aggregate current-noise term. When shot noise, dark-current noise, and thermal or electronic noise are enabled, its variance over detector bandwidth *B* is represented as

$$\sigma_{n,j}^2 = 2q\left(R_{\mathrm{pd}} P_j^{\mathrm{out}} + I_d\right)B + S_{\mathrm{el}} B$$

where *q* is the elementary charge, $I_d$ is the dark current, and $S_{el}$ is the input-referred electronic current-noise spectral density. Shot noise, dark-current noise, electronic noise, fanout loss, and imbalance are disabled in the reported experiments. The MZM and MRR implementations use the same ideal source distribution, responsivity scaling, numerical mapping, and calibration policy; only the programmed transfer element differs.

### B. MZM Weight Element

The MZM weight element is modeled as a balanced Mach–Zehnder interferometer in which the relative phase between the two arms controls the output optical power [3], [8]. For an ideal 50:50 input and output coupler, with insertion loss and finite extinction ratio included phenomenologically, the normalized power transmission is

$$T_{\mathrm{MZM}}(\Delta\phi) = \eta_{\mathrm{IL}} \left[T_{\min} + (1 - T_{\min}) \cos^2\left(\frac{\Delta\phi}{2}\right)\right]$$

where $T_{\mathrm{MZM}}$ is the optical-power transmission, $\Delta\phi$ is the differential phase between the interferometer arms, $\eta_{\mathrm{IL}} = 10^{-\mathrm{IL}/10}$ is the linear insertion-loss factor for insertion loss *IL* in decibels, and $T_{\min}$ is the finite transmission floor. When the extinction ratio *ER* is specified in decibels as the maximum-to-minimum transmitted-power ratio,

$$T_{\min} = 10^{-\mathrm{ER}/10}$$

The differential phase is decomposed into the calibrated bias, programmed control, and temperature-induced perturbation:

$$\Delta\phi = \phi_b + \phi_{\mathrm{ctrl}} + \Delta\phi_{\mathrm{T}}$$

For voltage control, the programmed phase is written as

$$\phi_{\mathrm{ctrl}} = \frac{\pi V}{V_\pi}$$

where $V$ is the differential control voltage and $V_\pi$ is the differential voltage required to produce a $\pi$-radian phase change under the adopted drive convention. If the implementation programs phase directly, $\phi_{\mathrm{ctrl}}$ is used without the voltage conversion.

For a uniform temperature offset $\Delta T$, the first-order differential thermal phase shift is

$$\Delta\phi_T = \kappa_{\phi,T}\Delta T$$

where $\kappa_{\phi,T} = \frac{\partial(\Delta\phi)}{\partial T}$ is the differential thermal phase coefficient. Equivalently, when the arm length $L$ and differential effective-index response are modeled explicitly,

$$\kappa_{\phi,T} = \frac{2\pi L}{\lambda_0}\frac{\partial(\Delta n_{\mathrm{eff}})}{\partial T}$$

with carrier wavelength $\lambda_0$ and arm effective-index difference $\Delta n_{\mathrm{eff}}$. A common-mode phase shift applied identically to both arms cancels from $\Delta\phi$; only a differential thermo-optic or path-length change alters the ideal MZM transmission.

Because $T_{\mathrm{MZM}}$ is nonnegative, signed neural-network weights are represented using the differential paths introduced in Section III-A. If $T_{ij,+}$ and $T_{ij,-}$ are the positive- and negative-path transmissions, the effective normalized weight is

$$w_{ij} = \frac{T_{ij,+} - T_{ij,-}}{\eta_{\mathrm{IL}}(1 - T_{\min})}$$

subject to the configured calibration and clipping range. The reported comparison uses the ideal calibrated MZM transfer, with no additional MZM-specific loss, extinction-ratio, bias, or thermal-coefficient variation enabled.

### C. MRR Weight Element

The MRR weight element is modeled as a single-bus all-pass microring, and the through-port optical-power transmission is controlled by the detuning between the optical carrier and a ring resonance [9], [10]. This transfer function applies only to the through port of a single-bus ring; an add-drop implementation requires the corresponding drop-port expression. With field self-coupling coefficient $r$, round-trip field transmission $a$, and round-trip phase $\theta$, the through-port power transmission is

$$T_{\mathrm{MRR}}(\theta) = \eta_{\mathrm{IL}}\frac{a^2 + r^2 - 2ar\cos(\theta)}{1 + a^2r^2 - 2ar\cos(\theta)}$$

where $\eta_{\mathrm{IL}} = 10^{-\mathrm{IL}/10}$ represents wavelength-independent insertion loss external to the resonant response. For a lossless directional coupler with real field coefficients, the cross-coupling coefficient $\kappa$ satisfies $r^2 + \kappa^2 = 1$. The round-trip field transmission is $a = \exp\left(-\frac{\alpha_p L_{\mathrm{rt}}}{2}\right)$, where $\alpha_p$ is the power-attenuation coefficient in inverse-length units and $L_{\mathrm{rt}} = 2\pi R$ is the ring circumference for radius $R$. Equivalently, if an amplitude-attenuation coefficient is used, its factor of two must not be applied again.

The round-trip phase at optical wavelength $\lambda$ and temperature $T$ is

$$\theta(\lambda, T) = \frac{2\pi n_{\text{eff}}(\lambda, T) L_{\text{rt}}}{\lambda}$$

and resonance occurs when $\theta(\lambda_r, T) = 2\pi m$ for integer mode number $m$. The programmed operating point is specified by the carrier-to-resonance detuning

$$\delta\lambda = \lambda_c - \lambda_r$$

where $\lambda_c$ is the optical-carrier wavelength. Near an isolated resonance, an equivalent loaded-quality-factor representation uses the full width at half maximum

$$\Delta\lambda_{\text{FWHM}} = \frac{\lambda_r}{Q_L}$$

where $Q_L$ is the loaded quality factor, including intrinsic and coupling losses. When intrinsic and coupling quality factors are used explicitly,

$$\frac{1}{Q_L} = \frac{1}{Q_i} + \frac{1}{Q_c}$$

A uniform temperature offset shifts the resonance relative to the fixed carrier. To first order,

$$\lambda_r(T_0 + \Delta T) = \lambda_{r,0} + \kappa_{\lambda,T} \Delta T$$

with thermal wavelength coefficient

$$\kappa_{\lambda,T} = \frac{\partial \lambda_r}{\partial T} \approx \frac{\lambda_{r,0}}{n_g} \left( \frac{\partial n_{\text{eff}}}{\partial T} + n_{\text{eff}} \alpha_L \right)$$

where $n_g$ is the group index and $\alpha_L$ is the linear thermal-expansion coefficient. This first-order expression assumes uniform ring temperature, small temperature excursions, and locally constant material and modal coefficients around $T_0$. The resulting temperature-dependent detuning is

$$\delta\lambda(\Delta T) = \delta\lambda_0 - \kappa_{\lambda,T} \Delta T$$

which is substituted into the resonant transfer function through $\theta(\lambda_c, T)$. Because $T_{MRR}$ is nonnegative, signed neural-network weights use the same differential-path convention as the MZM. If $T_{ij,+}$ and $T_{ij,-}$ are the calibrated positive- and negative-path transmissions, the signed weight is normalized by the calibrated differential-transmission span:

$$w_{ij} = \frac{T_{ij,+} - T_{ij,-}}{\Delta T_{\text{span}}}$$

where $\Delta T_{\text{span}}$ is the calibrated differential-transmission span used for weight normalization. This transmission span is dimensionless and is distinct from the temperature offset $\Delta T$. The calibration must ensure that $T_{ij,+}$ and $T_{ij,-}$ remain within the physically reachable through-port transmission range over the stated weight interval. The reported comparison uses the ideal calibrated MRR transfer, with no additional MRR-specific propagation-loss, coupling, linewidth, detuning, or thermal-coefficient variation enabled.

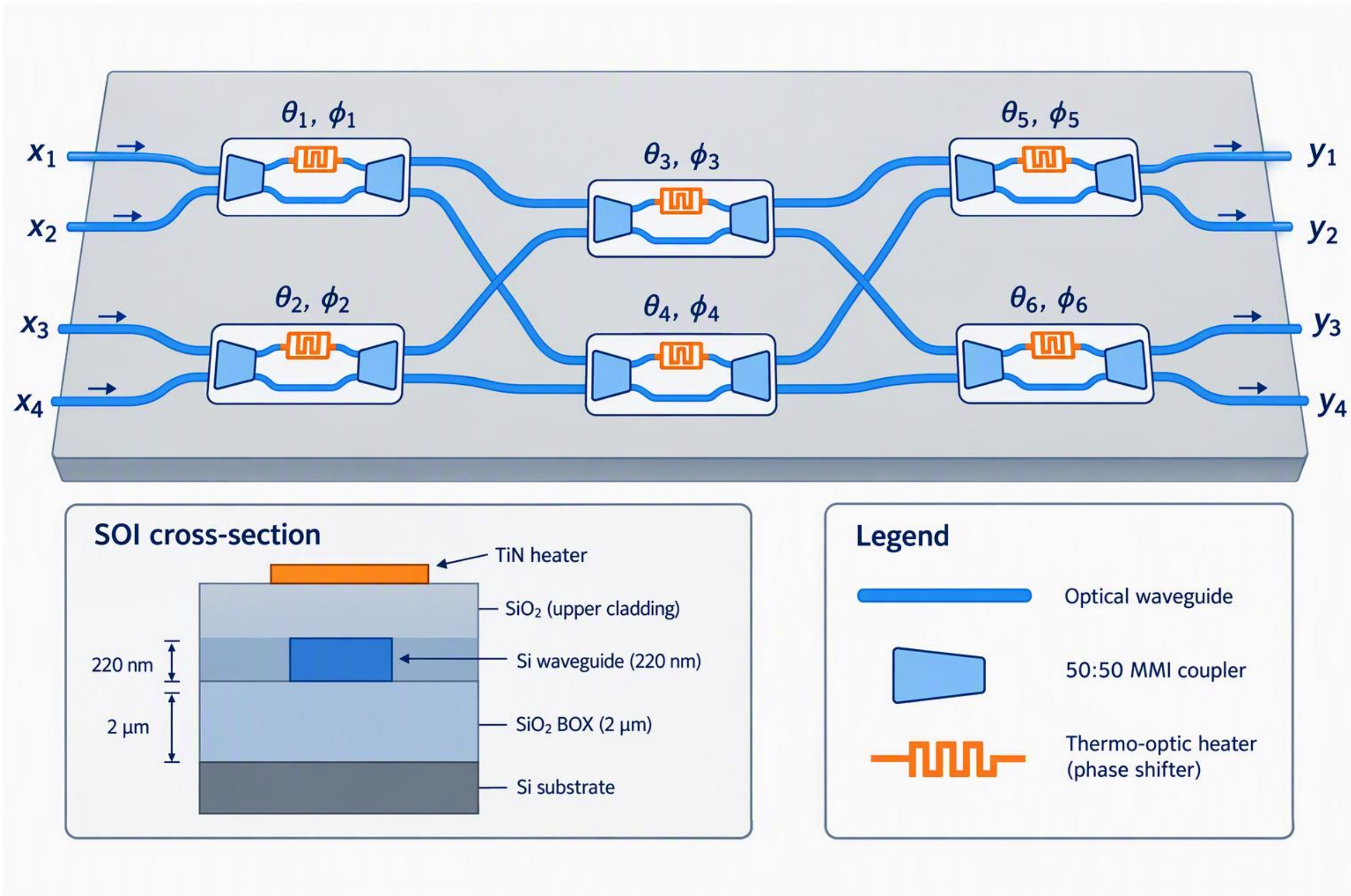

**Fig. 3. Mach–Zehnder modulator (MZM) weight element used in the photonic matrix multiplier.** The differential phase between the interferometer arms controls the optical-power transmission.

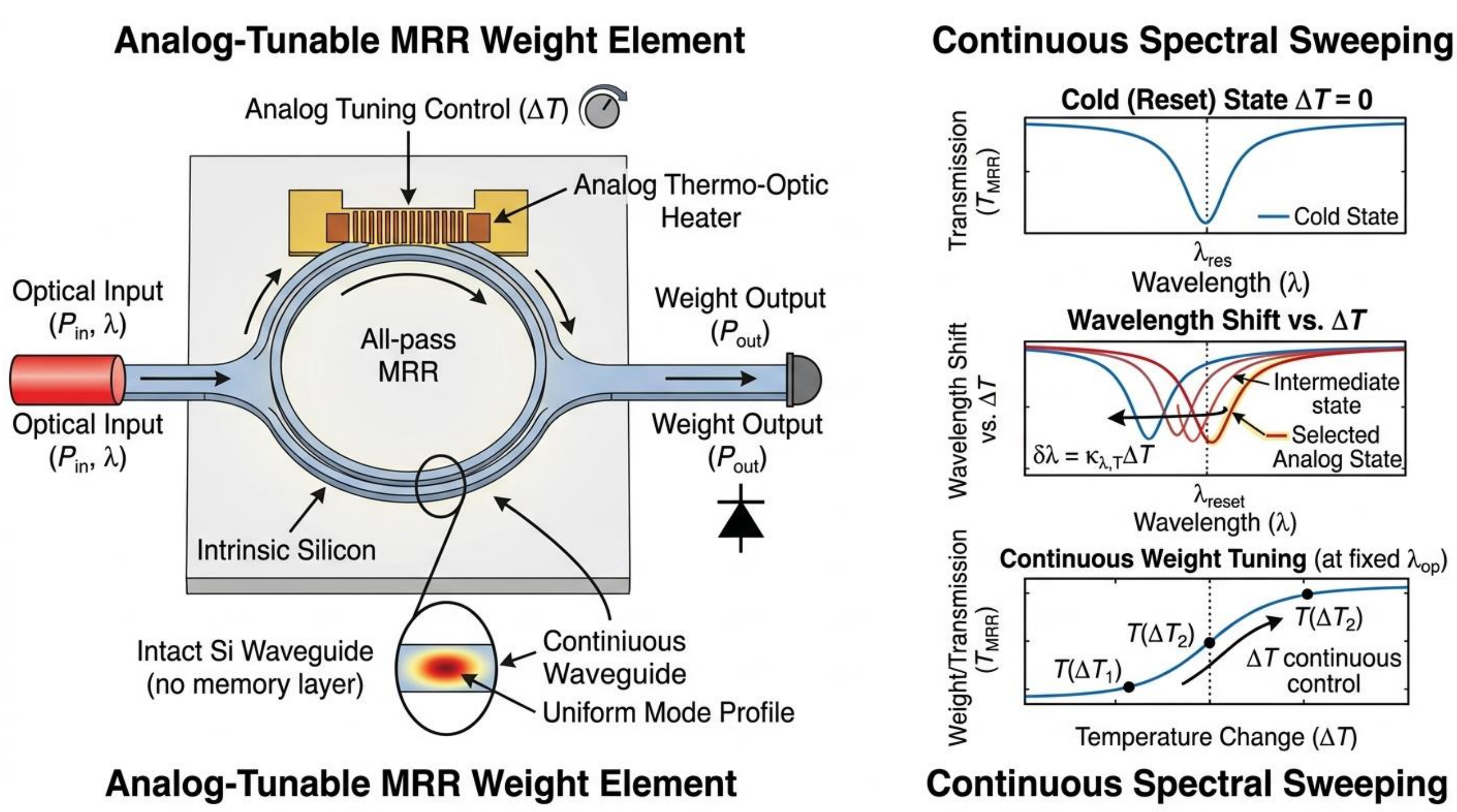

**Fig. 4. Schematic of an all-pass microring resonator (MRR) weight element operating under analog thermo-optic tuning.**

# IV. Experimental Methodology

## A. Laser Relative-Intensity Noise

Laser relative-intensity noise (RIN) is modeled as a zero-mean fractional fluctuation of the optical source power [11]. For source channel *i*,

$$P_{s,i}(t) = \overline{P}_{s,i}[1 + \varepsilon_i(t)]$$

where $\overline{P}_{s,i}$ is the mean source power and $\varepsilon_i(t)$ is the dimensionless fractional intensity fluctuation with $\mathbb{E}[\varepsilon_i(t)] = 0$. The one-sided RIN power spectral density is defined by

$$S_{\mathrm{RIN},i}(f) = \frac{S_{P,i}(f)}{\overline{P}_{s,i}^{\,2}}$$

where $S_{P,i}(f)$ is the one-sided power spectral density of the zero-mean fluctuation $\delta P_{s,i}(t) = P_{s,i}(t) - \overline{P}_{s,i}$ and therefore has units of $\mathrm{W}^2/\mathrm{Hz}$. Accordingly, $S_{\mathrm{RIN},i}(f)$ has units of $\mathrm{Hz}^{-1}$.

Its logarithmic representation is

$$\mathrm{RIN}_{\mathrm{dB/Hz}}(f) = 10\log_{10}\left(\frac{S_{\mathrm{RIN}}(f)}{1\ \mathrm{Hz}^{-1}}\right)$$

For an equivalent detection or sampling transfer function $H(f)$, the integrated fractional-power variance is

$$\sigma_{\mathrm{RIN},i}^2 = \int_0^\infty S_{\mathrm{RIN},i}\,(f)|H(f)|^2\,df$$

For spectrally white RIN with constant density $S_{\mathrm{RIN},i}$ over the relevant band, define the one-sided noise-equivalent bandwidth as

$$B_{\mathrm{NE}} = \frac{\int_0^\infty |H(f)|^2\ df}{|H(0)|^2}$$

If the transfer function is normalized to unity DC gain $|H(0)| = 1$, the integrated fractional-power variance reduces to

$$\sigma_{\mathrm{RIN},i}^2 = S_{\mathrm{RIN},i}B_{\mathrm{NE}}$$

and the corresponding root-mean-square fractional fluctuation is $\sigma_{\mathrm{RIN},i}$. Because multiple fanout branches may originate from the same laser, the shared source fluctuation is propagated to those branches as correlated multiplicative noise rather than regenerated independently at every weight element. For the fractional fluctuations $\varepsilon_\mathrm{m}$ and $\varepsilon_\mathrm{n}$ observed on fanout branches *m* and *n* at a common sampling instant,

$$\mathrm{Cov}[\varepsilon_m, \varepsilon_n] = c_{mn}\sigma_{\mathrm{RIN},m}\sigma_{\mathrm{RIN},n}$$

where $c_{mn}$ is the configured zero-lag correlation coefficient between branch fluctuations. Branches driven by the same laser have $c_{mn} = 1$ for the shared laser-RIN component, although branch-specific noise added after the split can reduce the correlation of the total fluctuations. Branches driven by independent lasers have $c_{mn} = 0$ only when the laser fluctuations are statistically independent and no common-path noise is present. The evaluated dimensionless RMS fractional intensity fluctuations are $\sigma_{\mathrm{RIN}} \in \{0.005,\ 0.01,\ 0.02,\ 0.05,\ 0.10, 0.15, 0.20\}$ corresponding to 0.5%, 1%, 2%, 5%, 10%, 15%, and 20% RMS fluctuations, respectively. Each run records the random seed and adopted correlation matrix.

### B. Thermal Drift

Thermal and phase drift is modeled at the calibrated weight level as a slow, per-output-column multiplicative perturbation. For each inference, the effective weight is

$$\widetilde{w_{ij}} = w_{ij}\left(1 + d_j\right)$$

where $w_{ij}$ is the calibrated weight from input $i$ to output $j$, $\widetilde{w_{ij}}$ is the drift-perturbed weight, and $d_j$ is a dimensionless random drift factor shared by all weights in output column $j$. The drift factor is drawn once per inference according to

$$d_j \sim \mathcal{N}\left(0, \sigma_{\text{drift}}^2\right)$$

where $\sigma_{\text{drift}}$ is the dimensionless standard deviation of the column-wise multiplicative drift. Thus, $\mathbb{E}\left[d_j\right] = 0$ and $\text{Var}\left[d_j\right] = \sigma_{\text{drift}}^2$. Because the same $d_j$ multiplies every $w_{ij}$ in column $j$, the perturbation coherently boosts or suppresses output $j$ until the next calibration or inference draw, rather than introducing independent weight noise.

Conditioned on $d_j$, the drift-induced weight error is

$$\delta w_{ij} = \widetilde{w}_{ij} - w_{ij} = w_{ij} d_j$$

with conditional variance

$$\text{Var}\left[\delta w_{ij} \mid w_{ij}\right] = w_{ij}^2 \sigma_{\text{drift}}^2$$

For mutually independent output-column draws, the covariance is

$$\text{Cov}\left[d_j, d_k\right] = \sigma_{\text{drift}}^2 \delta_{jk}$$

For the independent output-column draws used in the reported simulations, the covariance is given by the preceding expression, where the indicated symbol is the Kronecker delta. Laser RIN perturbs optical-source power, whereas the drift model acts directly on the calibrated weights and remains constant across all inputs contributing to the same output column during an inference.

The evaluated drift standard deviations are

$$\sigma_{\text{drift}} \in \{0.005,\ 0.01,\ 0.02,\ 0.05,\ 0.10, 0.15, 0.20\}$$

corresponding to 0.5%, 1%, 2%, 5%, 10%, 15%, and 20% RMS column-wise multiplicative drift, respectively. The same independent-column drift distribution and draw schedule are applied to the MZM and MRR mappings. The nominal weight calibration remains fixed during each inference, and each run records the random seed.

### C. Experimental Matrix and Reproducibility

The experiment matrix evaluates two neural-network models, two photonic weight elements, and two perturbation parameters. Nominal MZM and MRR operation denotes the corresponding calibrated device transfer defined in Section III with $\sigma_{\text{RIN}} = 0$ and $\sigma_{\text{drift}} = 0$; it is not a separate ideal-photonic reference. For each model–modulator pair, one scenario sweeps laser RIN with drift disabled, one scenario sweeps column-wise drift with RIN disabled, and one scenario evaluates the full joint RIN–drift grid. This produces eight one-parameter scenarios and four joint two-parameter scenarios, for 12 photonic evaluation scenarios in total. The float32 digital result is retained only as the software baseline. Classification accuracy and confusion matrices are evaluated across the configured perturbation sweeps. Simulations were performed using the QPI Labs Photonic Inference Simulator, developed by the author's affiliated organization; the public tutorial documents the simulator workflow and available controls [13].

Reproducibility is maintained through the recorded model identifiers, modulator configuration, random seeds, experiment configurations, and raw result files. Table III enumerates the complete scenario matrix.

**TABLE III. TWELVE-SCENARIO EXPERIMENT MATRIX**

| Scenario | Model | Modulator | Evaluation | $\sigma_{RIN}$ | $\sigma_{drift}$ |
|---|---|---|---|---|---|
| **1** | Small CNN | MZM | RIN-only sweep | {0, 0.005, 0.01, 0.02, 0.05, 0.10, 0.15, 0.20} | 0 |
| **2** | Small CNN | MRR | RIN-only sweep | {0, 0.005, 0.01, 0.02, 0.05, 0.10, 0.15, 0.20} | 0 |
| **3** | ResNet-8 | MZM | RIN-only sweep | {0, 0.005, 0.01, 0.02, 0.05, 0.10, 0.15, 0.20} | 0 |
| **4** | ResNet-8 | MRR | RIN-only sweep | {0, 0.005, 0.01, 0.02, 0.05, 0.10, 0.15, 0.20} | 0 |
| **5** | Small CNN | MZM | Drift-only sweep | 0 | {0, 0.005, 0.01, 0.02, 0.05, 0.10, 0.15, 0.20} |
| **6** | Small CNN | MRR | Drift-only sweep | 0 | {0, 0.005, 0.01, 0.02, 0.05, 0.10, 0.15, 0.20} |
| **7** | ResNet-8 | MZM | Drift-only sweep | 0 | {0, 0.005, 0.01, 0.02, 0.05, 0.10, 0.15, 0.20} |
| **8** | ResNet-8 | MRR | Drift-only sweep | 0 | {0, 0.005, 0.01, 0.02, 0.05, 0.10, 0.15, 0.20} |
| **9** | Small CNN | MZM | Joint RIN–drift grid | Full set | Full set |
| **10** | Small CNN | MRR | Joint RIN–drift grid | Full set | Full set |
| **11** | ResNet-8 | MZM | Joint RIN–drift grid | Full set | Full set |
| **12** | ResNet-8 | MRR | Joint RIN–drift grid | Full set | Full set |

**Baseline condition: For each network, the point $\sigma_{RIN} = \sigma_{drift} = 0$ reproduces the same baseline accuracy for the digital, calibrated MZM, and calibrated MRR evaluations: 79.5% for the Small CNN and 70.0% for ResNet-8. This shared zero-perturbation point is not counted as an** **additional scenario; the 12 scenarios comprise the RIN-only, drift-only, and joint perturbation evaluations.**

## V. Results

The results follow the 12-scenario matrix in Table III. Table IV reports the common zero-perturbation baseline for each network; the digital, calibrated MZM, and calibrated MRR evaluations coincide at this point. Figs. 5 and 6 report the workload-specific baseline confusion matrices, and Figs. 7 and 8 report the corresponding rolling-accuracy traces. Figs. 9 and 10 report the four RIN-only scenarios, with scenarios 1 and 2 combined in Fig. 9 and scenarios 3 and 4 combined in Fig. 10. Figs. 11 and 12 report the four drift-only scenarios, with scenarios 5 and 6 combined in Fig. 11 and scenarios 7 and 8 combined in Fig. 12. Figs. 13–16 report the four joint RIN–drift scenarios individually.

### A. Common Zero-Perturbation Baseline

TABLE IV. COMMON ZERO-PERTURBATION BASELINES

| Network | Baseline accuracy (%) |
|---|---|
| **Small CNN** | 79.5 |
| **ResNet-8** | 70.0 |

**The baseline is shared by the digital, calibrated MZM, and calibrated MRR evaluations at $\sigma_{RIN} = \sigma_{drift} = 0$.**

The zero-perturbation baseline is reported using one class-normalized confusion matrix and one rolling-accuracy trace for each workload. The Small-CNN confusion matrix shows the distribution of correct predictions and class-specific confusions at the 79.5% baseline, while the ResNet-8 confusion matrix provides the corresponding classwise view at the 70.0% baseline. The rolling-accuracy plots show the cumulative test accuracy as successive samples are evaluated. These four baseline figures are common to the digital, calibrated MZM, and calibrated MRR evaluations because all three coincide when $\sigma_{RIN} = \sigma_{drift} = 0$.

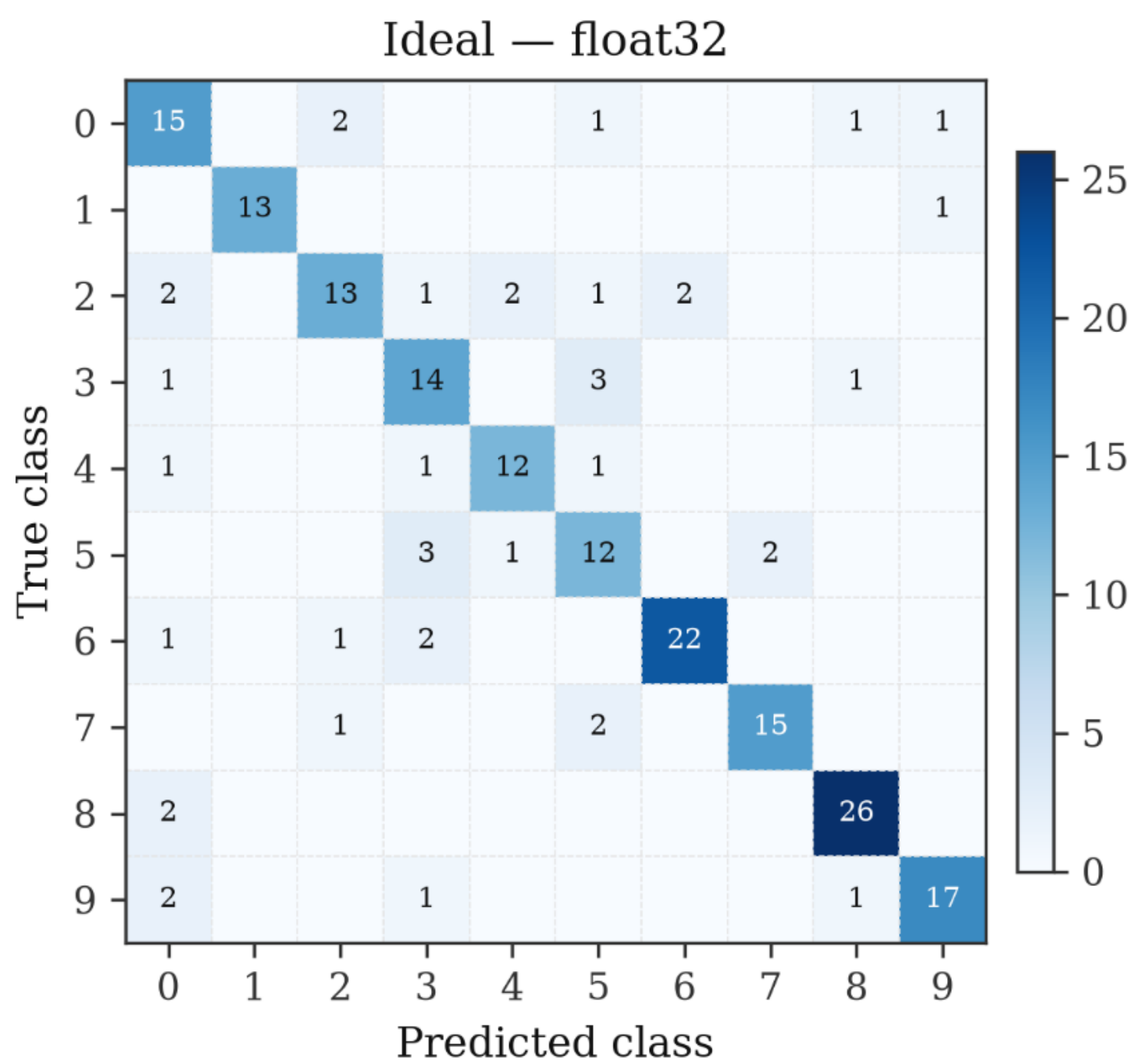

**Fig. 5. Class-normalized confusion matrix for the Small-CNN zero-perturbation baseline.**

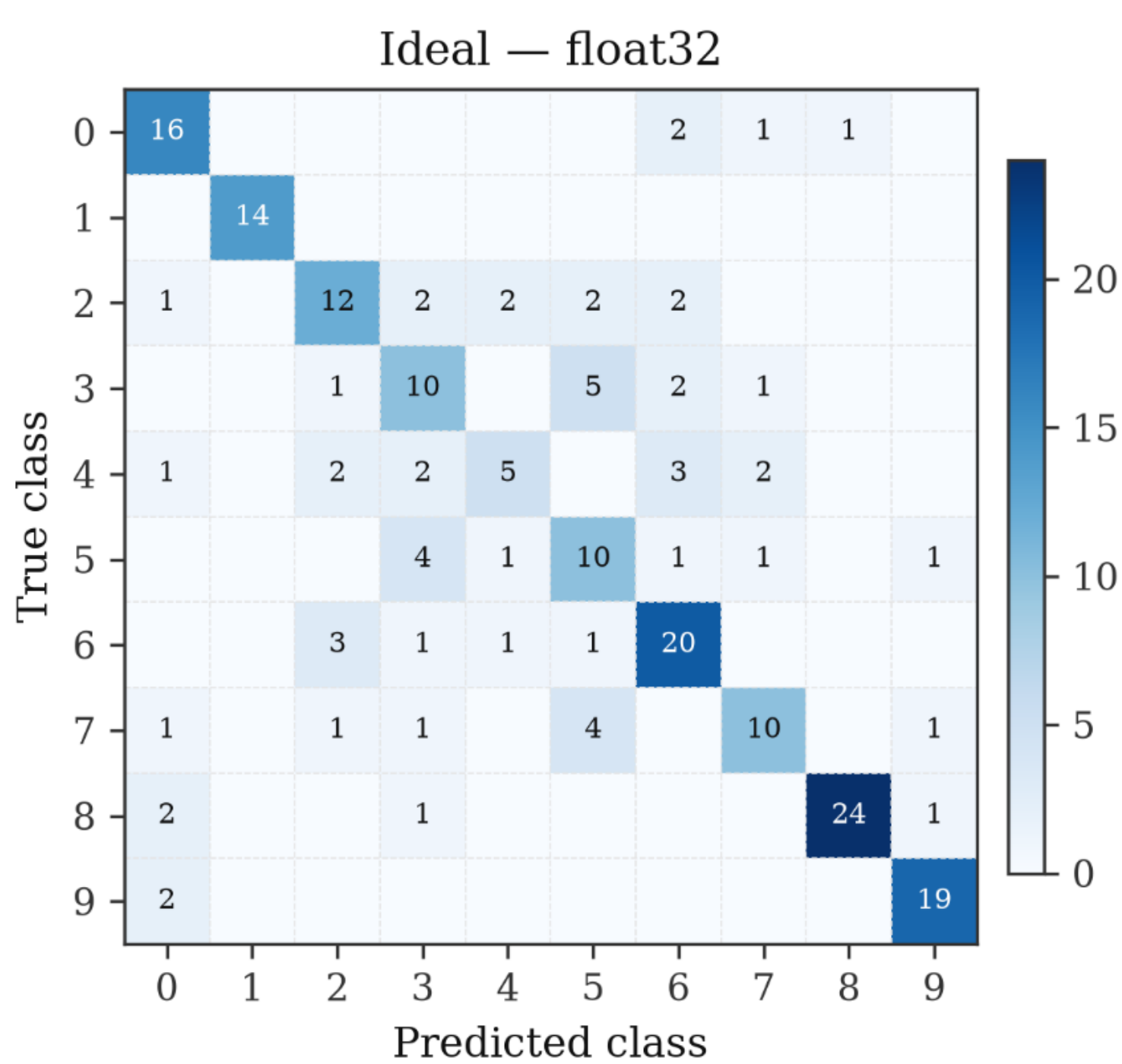

**Fig. 6. Class-normalized confusion matrix for the ResNet-8 zero-perturbation baseline.**

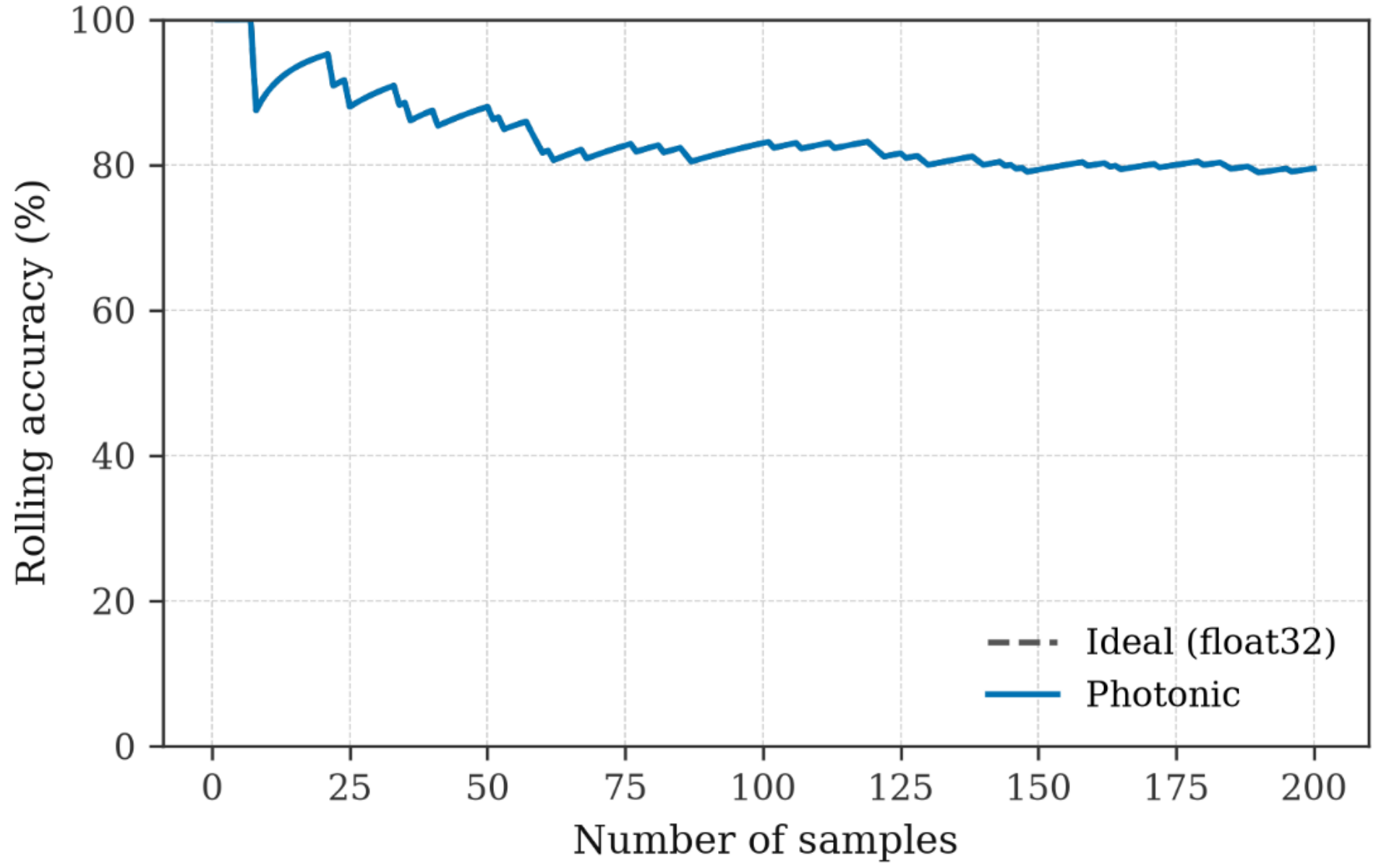

**Fig. 7. Rolling test accuracy for the Small-CNN zero-perturbation baseline.**

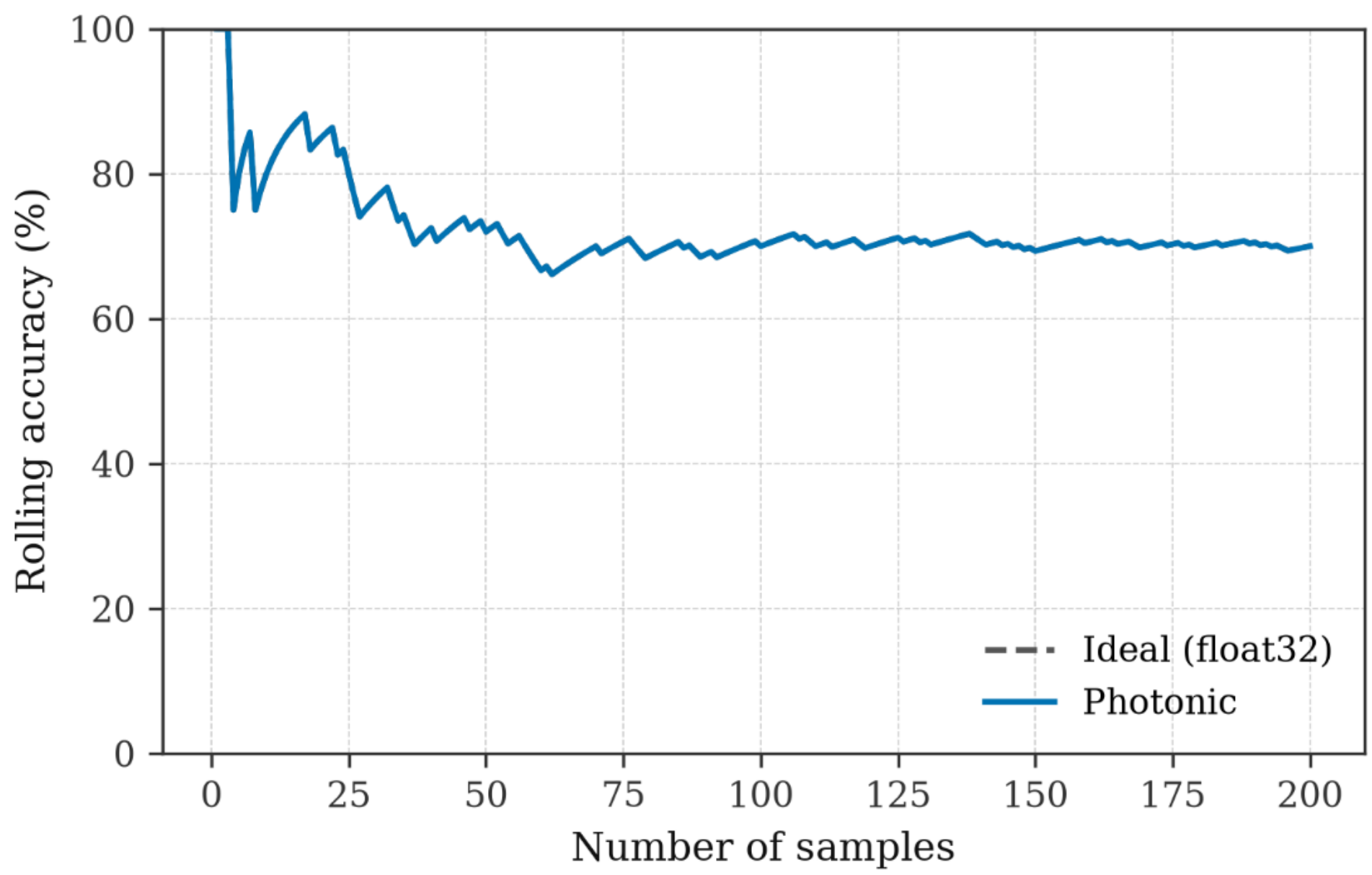

**Fig. 8. Rolling test accuracy for the ResNet-8 zero-perturbation baseline.**

### B. One-Dimensional Robustness Sweeps

The RIN-only results are reported for the four model–modulator pairs in scenarios 1–4. Fig. 9 compares the Small-CNN MZM and MRR results, and Fig. 10 compares the ResNet-8 MZM and MRR results. Each figure plots classification accuracy over the common $\sigma_{\mathrm{RIN}}$ sweep with $\sigma_{\mathrm{drift}} = 0$ and begins at the corresponding shared workload baseline. The paired presentation permits direct comparison of the two modulators within each workload.

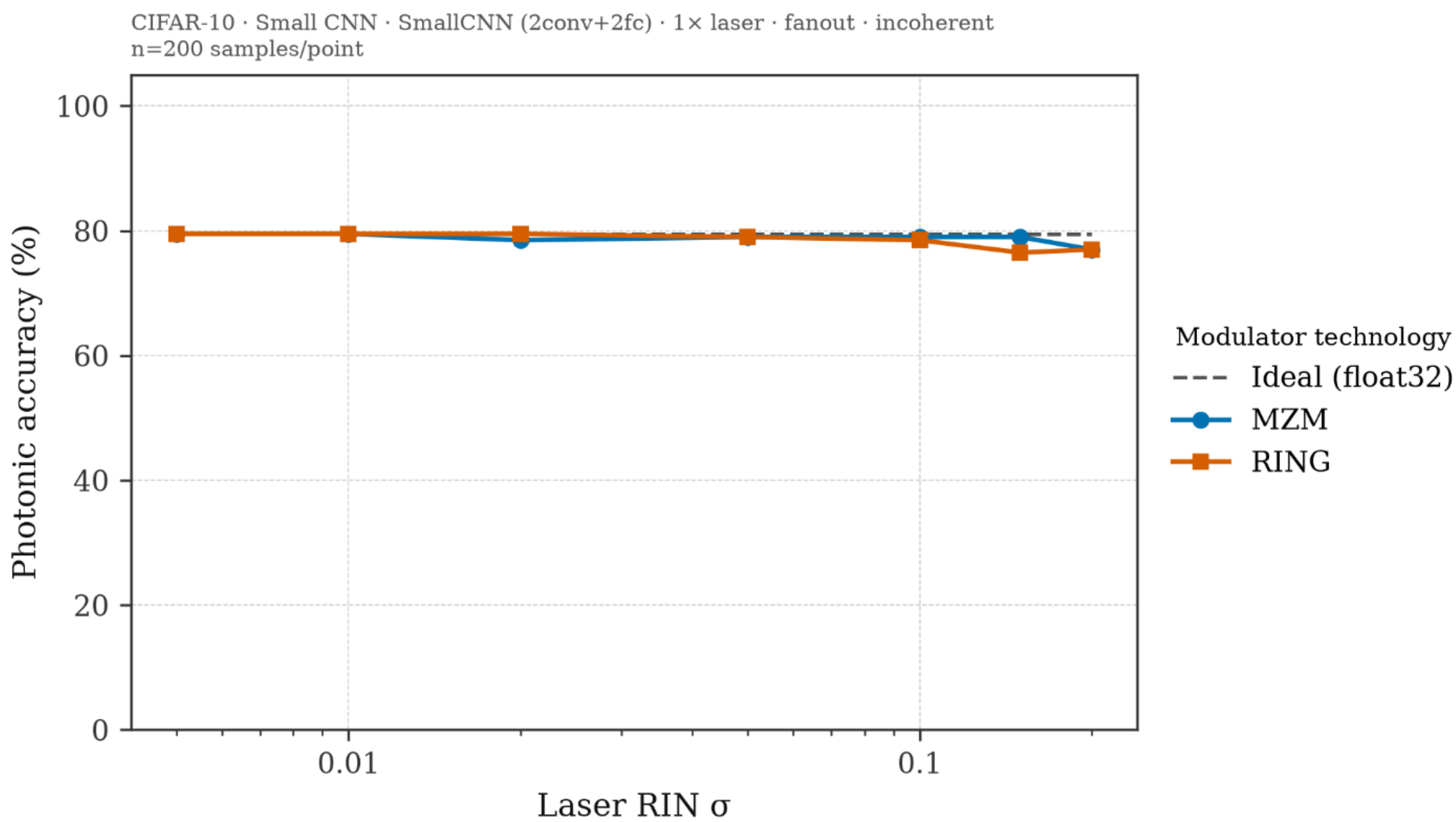

**Fig. 9. Small-CNN accuracy under the MZM and MRR RIN-only sweeps (scenarios 1 and 2).**

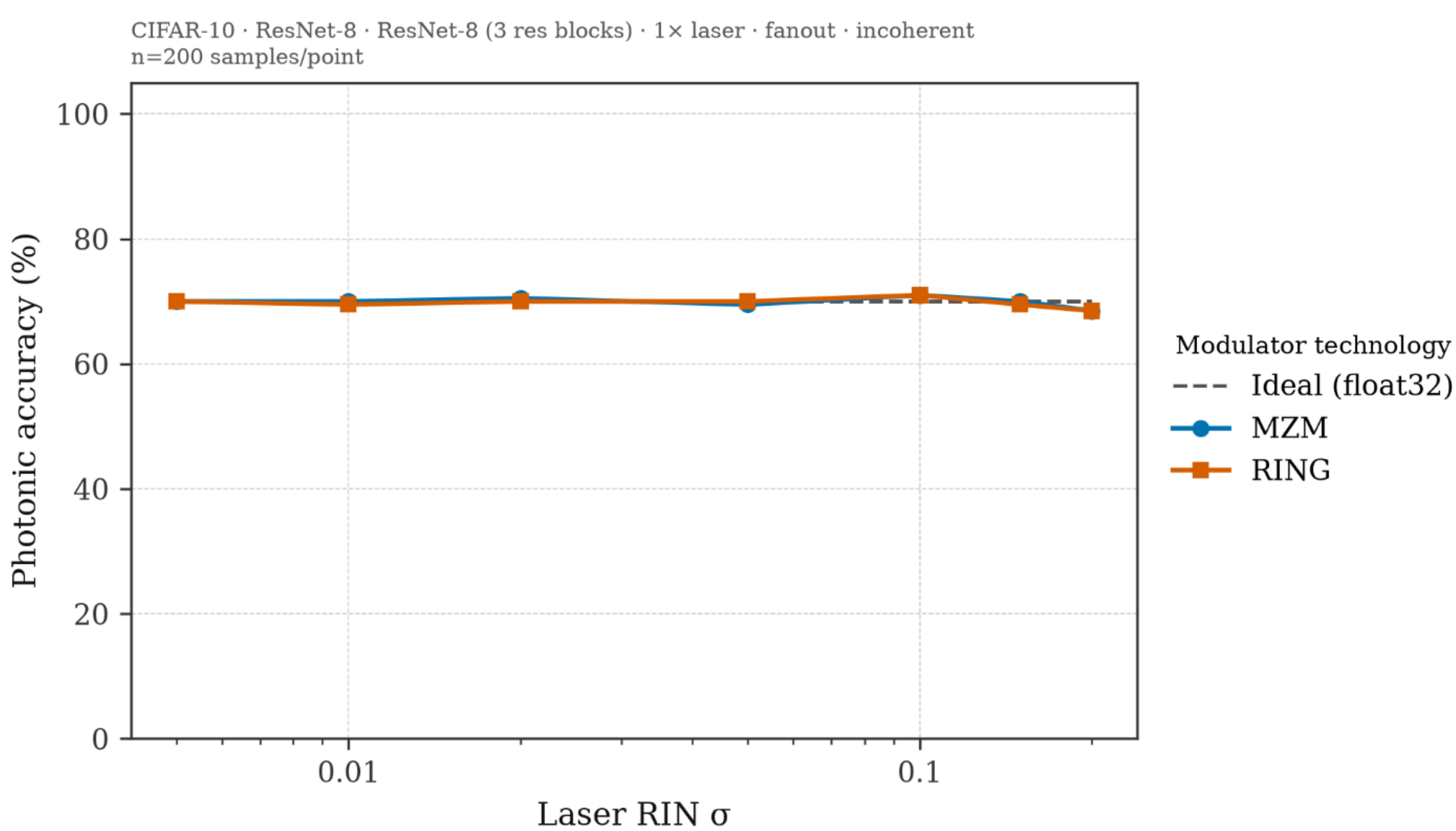

**Fig. 10. ResNet-8 accuracy under the MZM and MRR RIN-only sweeps (scenarios 3 and 4).**

The drift-only results are reported for the four model–modulator pairs in scenarios 5–8. Fig. 11 compares the Small-CNN MZM and MRR results, and Fig. 12 compares the ResNet-8 MZM and MRR results. Each figure plots classification accuracy over the common $\sigma_{\text{drift}}$ sweep with $\sigma_{\text{RIN}} = 0$ and begins at the corresponding shared workload baseline. The paired presentation permits direct comparison of the two modulators within each workload.

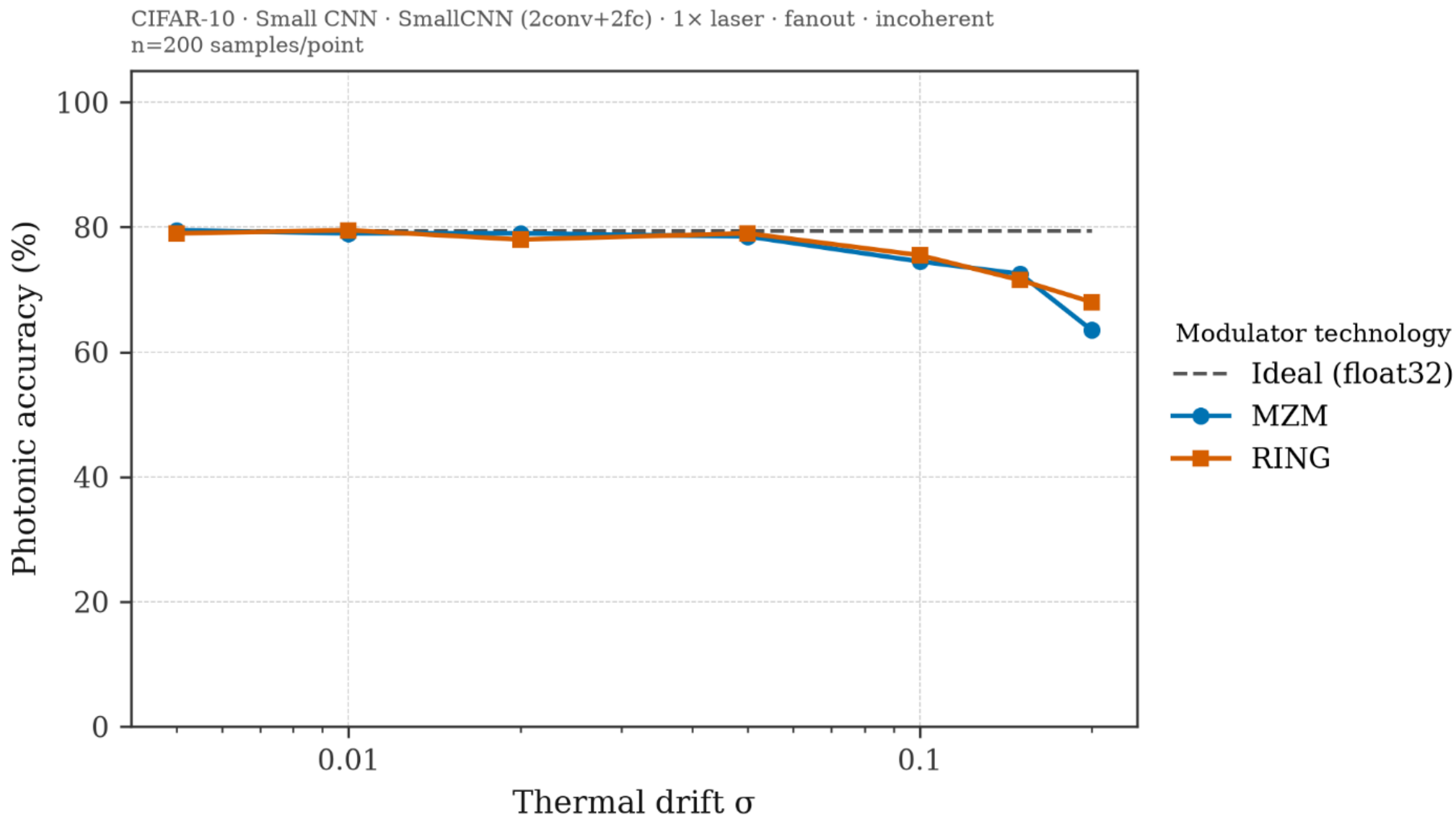

**Fig. 11. Small-CNN accuracy under the MZM and MRR drift-only sweeps (scenarios 5 and 6).**

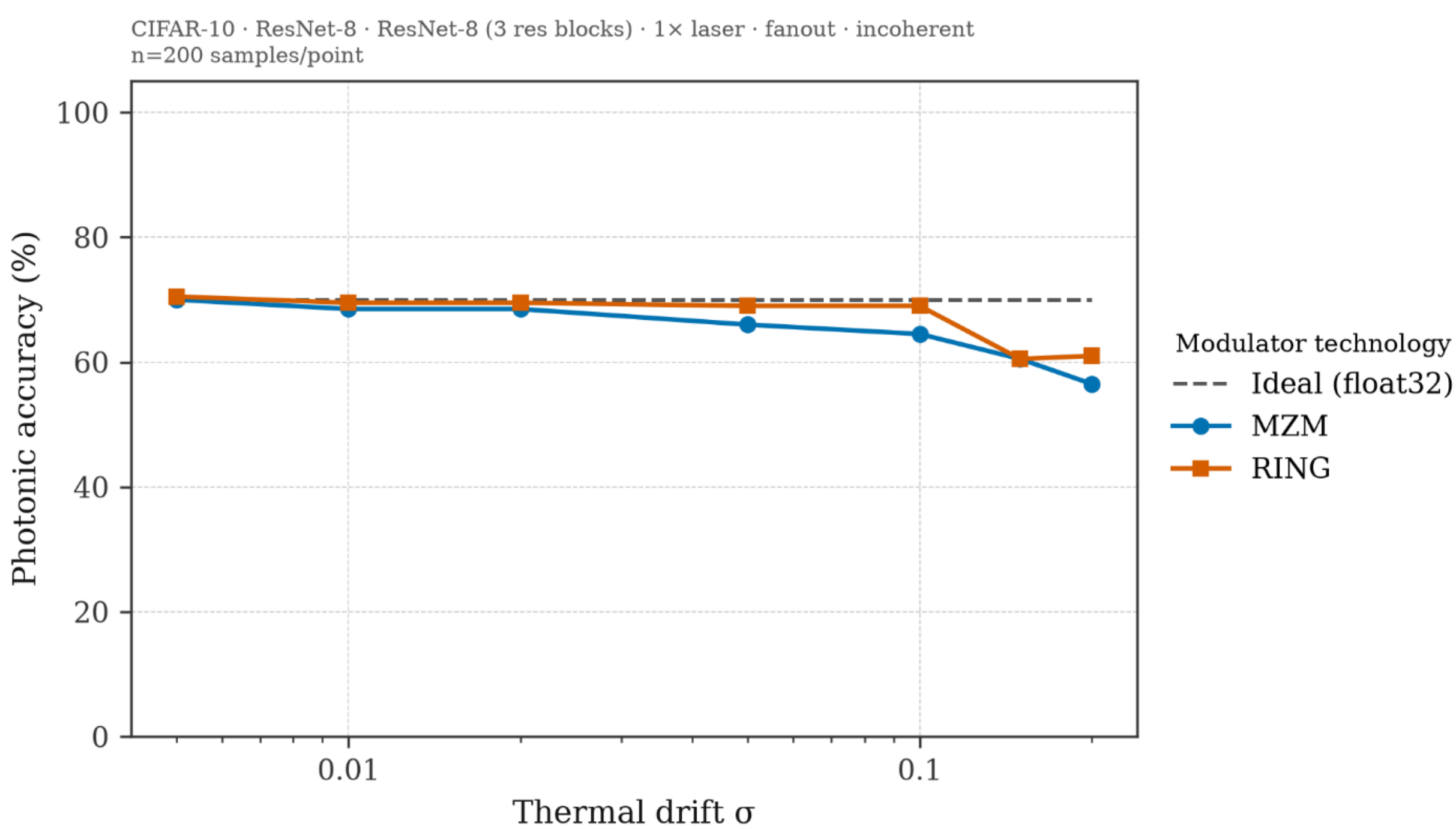

**Fig. 12. ResNet-8 accuracy under the MZM and MRR drift-only sweeps (scenarios 7 and 8).**

### C. Joint RIN–Drift Response

The joint RIN–drift response is reported separately for the four model–modulator pairs in scenarios 9–12. Each figure presents classification accuracy over the full $\sigma_{RIN} \times \sigma_{drift}$ grid, uses the corresponding common network baseline as the zero-loss reference, and marks the shared baseline point at (0, 0). The four figures use common axes and color scales to show the accuracy response to the combined perturbations for each model–modulator pair.

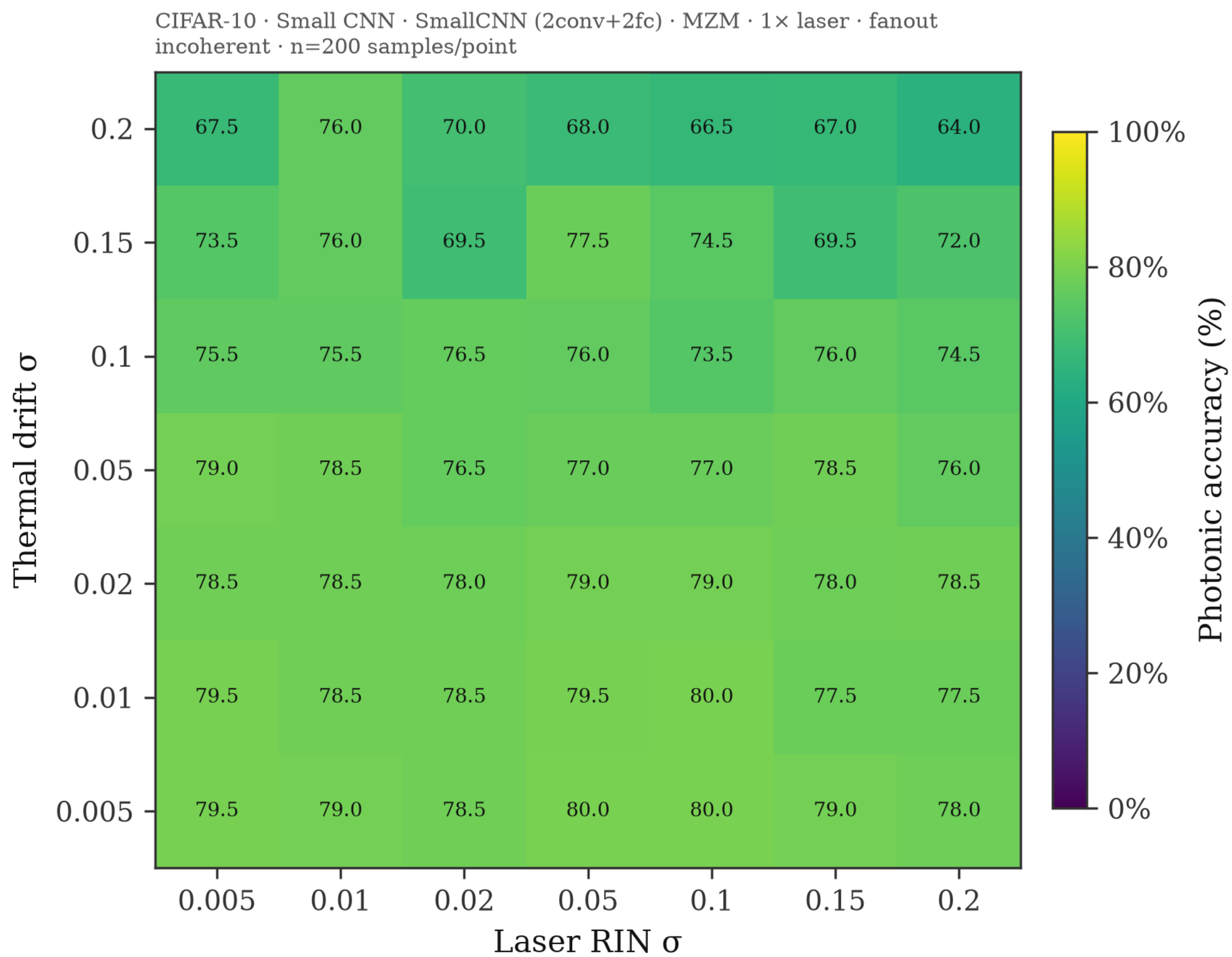

**Fig. 13. Joint RIN–drift accuracy for Small-CNN with MZM (scenario 9).**

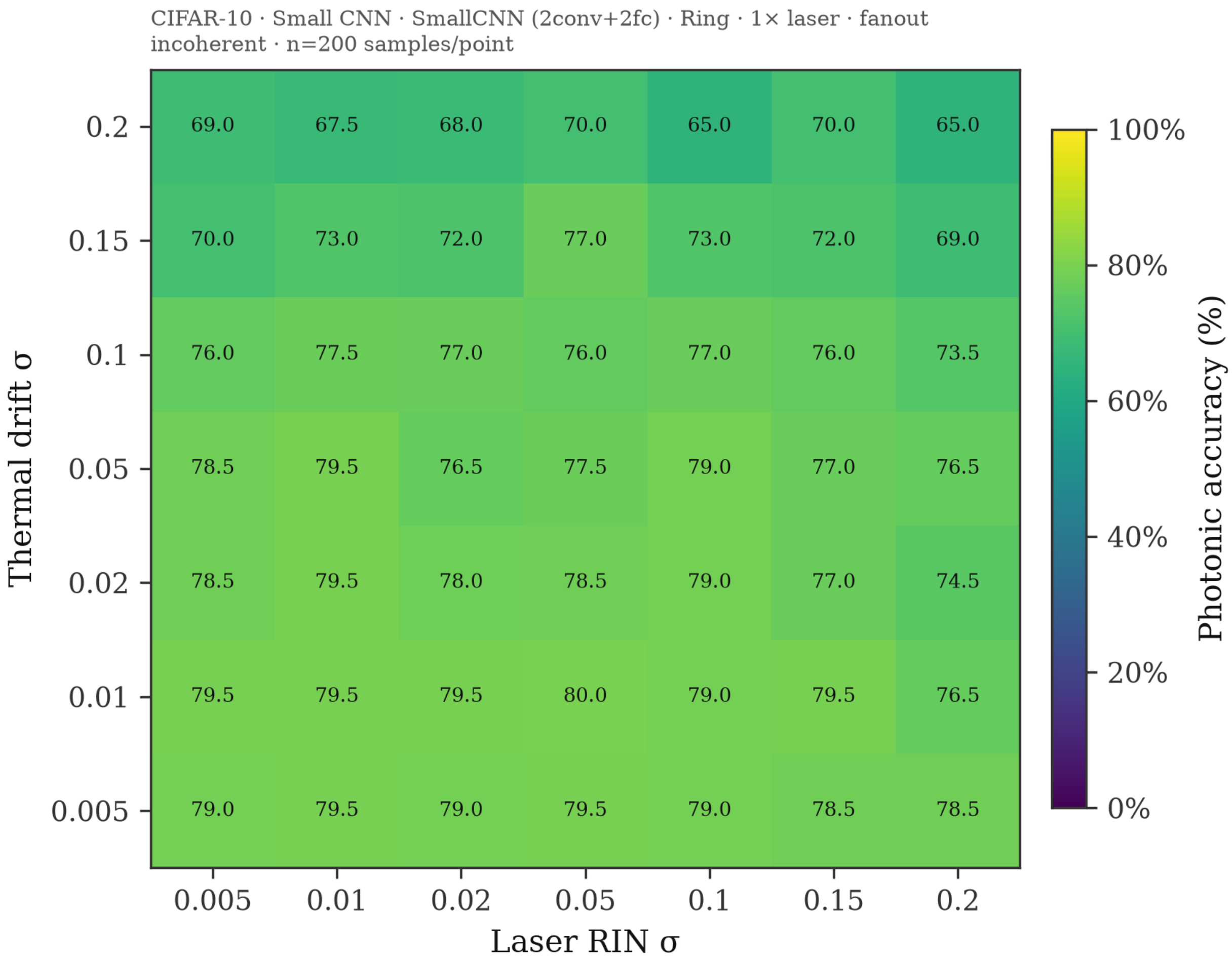

**Fig. 14. Joint RIN–drift accuracy for Small-CNN with MRR (scenario 10).**

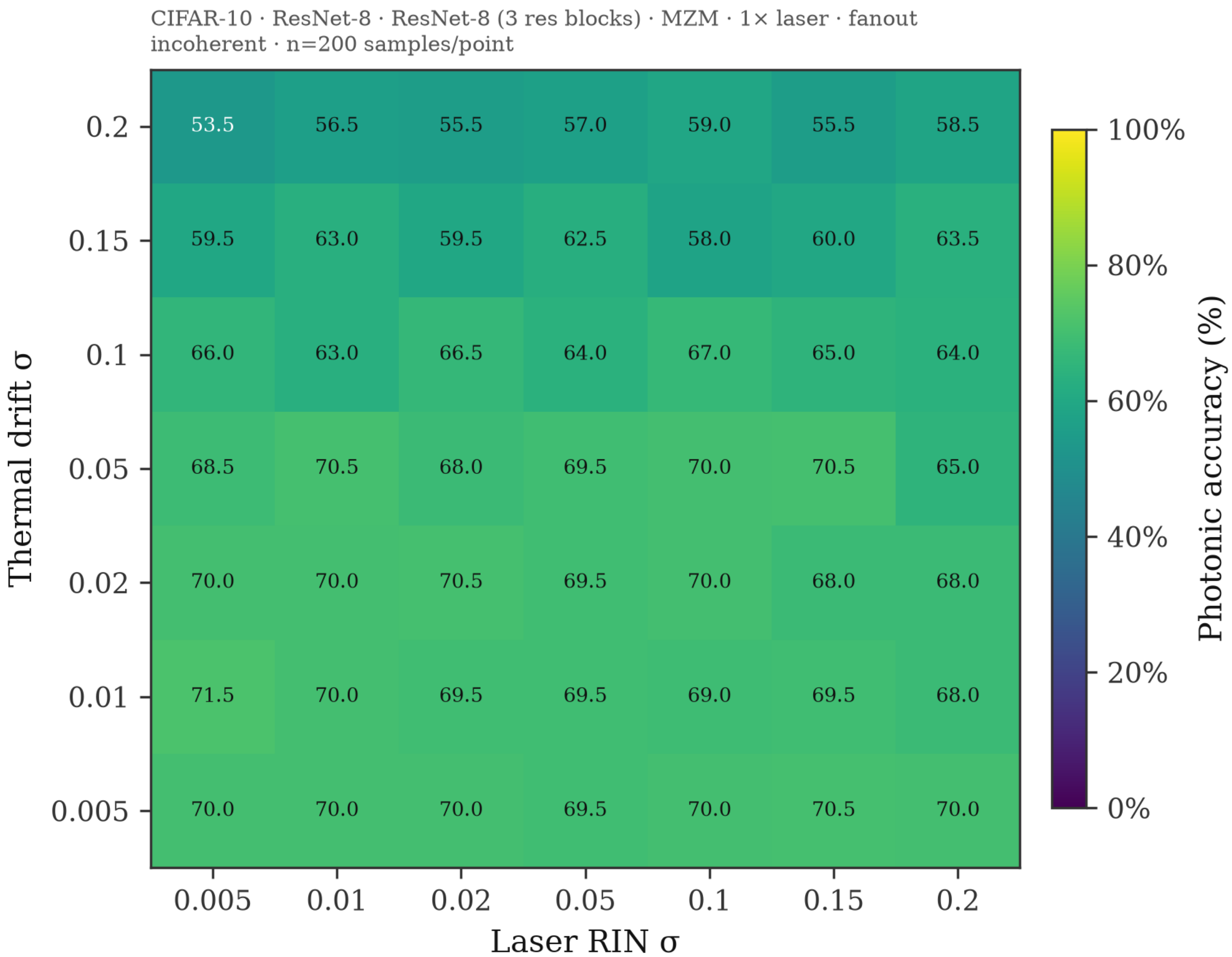

**Fig. 15. Joint RIN–drift accuracy for ResNet-8 with MZM (scenario 11).**

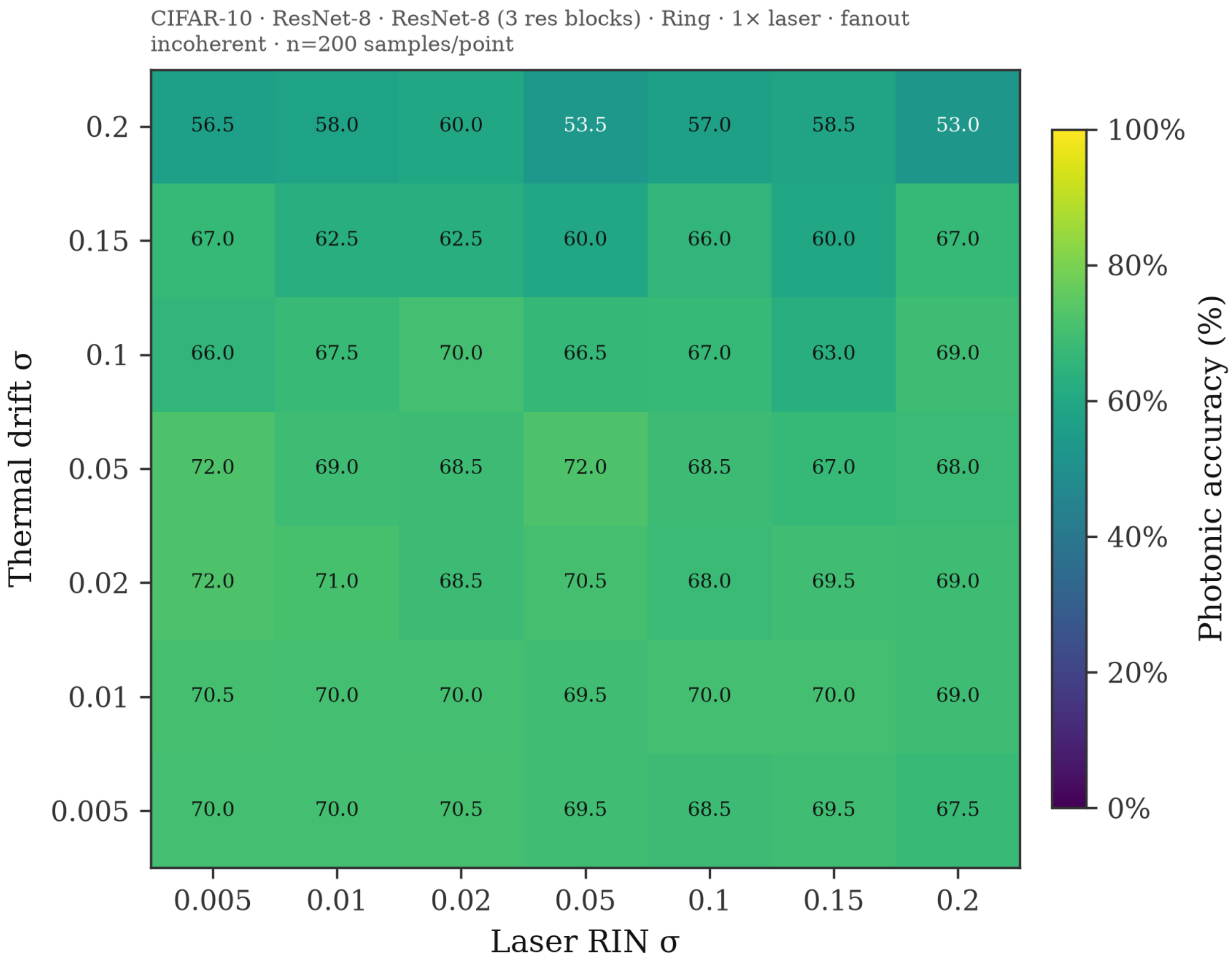

**Fig. 16. Joint RIN–drift accuracy for ResNet-8 with MRR (scenario 12).**

## VI. Discussion

The results establish a common reference point and then show how inference accuracy changes as optical-source noise and calibrated-weight drift are introduced. At zero perturbation, the digital, calibrated MZM, and calibrated MRR evaluations reproduce the same workload baselines: 79.5% for the Small CNN and 70.0% for ResNet-8. The confusion matrices and rolling-accuracy traces confirm the classwise and samplewise behavior of these reference configurations. The RIN-only and drift-only sweeps then separate the contribution of each nonideality, while the joint response surfaces identify the combined operating regions over the full evaluated perturbation grid. Taken together, these results show that robustness is not determined by the nominal weight element alone; it also depends on the network mapping, perturbation magnitude, signed-weight representation, calibration, and accumulation path.

The principal device distinction is the interferometric MZM response versus the resonant MRR response; however, the reported drift model is deliberately applied at the calibrated-weight level as a common per-output-column multiplicative factor for both architectures. Likewise, source-noise correlation follows the configured source and distribution topology rather than being regenerated independently at every multiply. The paired MZM–MRR comparisons therefore evaluate complete mapped inference chains under matched assumptions, not isolated device transfer curves. Within this controlled framework, the simulated accuracy curves and joint response surfaces contain operating configurations for which both photonic implementations retain useful AI inference performance. This finding supports the feasibility of photonic matrix multiplication for the evaluated AI workloads when source stability, drift, calibration, and operating point are selected together. It does not imply that every point in the evaluated grid is acceptable, nor does it establish a universal tolerance threshold independent of workload or implementation.

## VII. Limitations

The simulations isolate laser RIN and column-wise thermal drift in the evaluated photonic inference configurations. They do not model the full set of effects that may determine fabricated-system performance, including device-to-device variation, fabrication yield, wavelength-dependent loss, nonlinear detector or electronic behavior, packaging parasitics, long-term aging, control-loop dynamics, recalibration latency, and stabilization power. The drift abstraction acts on calibrated weights rather than deriving every perturbation from a spatially resolved thermal model, and the reported RIN levels depend on the adopted bandwidth, sampling, and correlation assumptions. Quantitative conclusions therefore remain specific to the selected device models, parameter ranges, calibration policy, numerical precision, layer mapping, CIFAR-10 dataset, and two network families. The identified feasible operating configurations should consequently be interpreted as simulation-supported design regions that motivate hardware validation, not as guaranteed specifications for a fabricated accelerator. Future work should include measured device distributions, closed-loop control overhead, broader datasets and architectures, repeated trials across seeds, and experimental comparison with physical MZM- and MRR-based arrays.

## VIII. Conclusion

This paper compared MZM- and MRR-based photonic matrix multipliers under matched source, distribution, detection, calibration, and numerical assumptions for a compact CIFAR-10 CNN and ResNet-8. Both photonic mappings reproduce the digital zero-perturbation baselines of 79.5% and 70.0%, respectively. Separate RIN and drift sweeps reveal the sensitivity to each perturbation, and the joint response surfaces show how their combined action defines workload- and architecture-dependent operating regions. The simulation results demonstrate that the evaluated design space includes configurations in which photonic matrix multiplication can support useful AI inference performance. Feasibility therefore depends on co-selecting the photonic weight element, source-noise level, drift tolerance, calibration strategy, and neural-network mapping rather than treating any one device parameter in isolation. These results provide system-level evidence that photonic compute is a viable candidate for the evaluated AI workloads under controlled operating conditions, while stopping short of claiming fabricated-hardware performance. Hardware measurements and broader workload studies are the next steps needed to translate the identified simulation regions into implementation specifications.